# Decision-support strategies for photovoltaic self-consumption under declining electricity prices and limited remuneration of surplus generation.

Ana B. Cristóbal*[a,b] (0000-0002-4314-6160), Daniel Sierra[a], Laura Palomino[b], Luis Miguel Carrasco [a,b], Luis Narvarte (0000-0002-6289-7605)[b].

a) Escuela Técnica Superior de Ingeniería de Sistemas Informáticos, Universidad Politécnica de Madrid, C/Alan Turing s/n, 28031 Madrid.

b) Instituto de Energía Solar, Universidad Politécnica de Madrid, C/Nikola Tesla s/n, 28031 Madrid, Spain.

anabelen.cristobal@upm.es, daniel.sierra@alumnos.upm.es, laura.palomino@upm.es, luis.narvarte@upm.es

## Abstract

The success of distributed photovoltaics may be undermining its own future. As solar penetration increases, electricity prices fall during periods of peak generation, reducing the value of surplus photovoltaic production. This raises a critical question for the energy transition: can citizen-led energy systems remain economically viable in electricity markets increasingly dominated by renewable generation?

Rather than exploring technically optimal but institutionally unrealistic solutions, we investigate what options remain available to ordinary citizens under current regulatory and market conditions. Using high-resolution consumption data of a mature rural community sharing a PV facility among 24 users, providing evidence-based insights into pathways for long-term sustainability. Key contributions are advanced. First, the analysis redefines the value proposition of collective energy systems, showing that robust internal coordination can mobilize participation and investment as effectively as external subsidies, an insight particularly relevant for contexts facing capital constraints. Second, the study evaluates alternative energy-sharing mechanisms that move beyond fixed equity-based allocations. By systematically comparing static, dynamic, and hybrid models—with and without storage integration—it develops an adaptable framework that balances fairness and flexibility while aligning with evolving governance needs.

Results show that collective self-consumption generates substantial coordination benefits, reducing required PV capacity, lowering investment costs, and increasing annual savings compared with equivalent individually operated systems. Alternative allocation schemes further improve the distribution of benefits among participants and enhance the utilization of locally generated electricity, although gains depend on the trade-off between efficiency, fairness, and governance complexity. By contrast, under current electricity prices and remuneration schemes, battery storage provides only limited additional economic value, becoming attractive only under specific market conditions. The findings suggest that the long-term viability of citizen-led photovoltaic initiatives may depend less on technological sophistication than on their ability to leverage collective coordination and adaptive governance arrangements.

## 1. Introduction

The transition towards a more sustainable, decentralized, and equitable energy system has gained momentum worldwide in response to the combined challenges of climate change, energy security, and social justice. Among the different technologies supporting this transition, distributed photovoltaic generation has become one of the most widely adopted solutions due to its technological maturity and declining installation costs.

Paradoxically, this success is creating new challenges. As solar penetration increases, electricity prices during periods of peak PV generation tend to decline, reducing the value of surplus electricity exported to the grid. In many countries, exported electricity receives limited compensation, meaning that a growing share of PV generation is produced precisely when its economic value is lowest[1]. This raises a critical question for the energy transition: can citizen-led energy systems remain economically viable in electricity markets increasingly dominated by renewable generation?

Under these conditions, maximizing the local utilization of renewable electricity becomes increasingly important for maintaining the attractiveness of PV self-consumption.

For individual photovoltaic installations, achieving high self-consumption rates remains challenging because household demand profiles rarely coincide with peak solar production and alternatives such as batteries or heat-pumps requests higher investments and are not always feasible. This situation raises important questions regarding the long-term profitability of purely individual self-consumption models and highlights the need for alternative approaches capable of increasing local utilization of renewable generation and keep citizens engaged with the energy transition.

In parallel, energy communities have gained increasing attention across Europe as a novel organizational model for citizen participation in the energy transition[2]. Supported by European regulatory frameworks, these initiatives enable citizens to collectively engage in the generation, management, and consumption of renewable energy. Beyond their contribution to decarbonization, energy communities have been associated with broader objectives including energy democracy, citizen empowerment, local value creation, and social cohesion [3,4].

In Spain, this momentum has materialized in numerous citizen-led projects[5] that, supported primarily by European regulatory frameworks [6,7], have begun to develop models of shared generation, collective self-consumption, and participatory governance. These citizen initiatives are paving a new path toward models that, beyond technical and economic design, generate social returns aligned with local needs. In practice, however, regulatory and administrative complexity has limited the development of most energy communities to photovoltaic collective generation and self-consumption schemes. [8]

Many citizen-led photovoltaic energy projects -residential and collective- have been established during periods of higher electricity prices and strong public support schemes.

However, recent market developments, including declining electricity prices, limited remuneration for surplus energy exports, and the gradual reduction of public support schemes, raise important questions about the long-term economic viability of these initiatives and the role that policymakers expect citizens to play in the energy transition. As example, in Spain, photovoltaic self-consumption continues to expand and has already surpassed 9.3 GW of cumulative installed capacity. However, the annual deployment rate has been slowing down for several years. In 2025, 1,139 MW of new self-consumption capacity were installed, representing a 3.7% decrease compared to 2024 and remaining well below the peak reached in 2022. The slowdown was particularly pronounced in the residential sector, where new installations amounted to 229 MW, representing a 17% year-on-year decline[9]. These figures suggest a transition from a phase of rapid expansion to a more mature market, where maximizing the value of existing installations is becoming increasingly important and where new approaches are needed to support citizen participation to local generation and consumption beyond direct subsidies.

Under these changing conditions, understanding how citizen-led projects can remain attractive and continue delivering value to their members becomes increasingly relevant.

While significant advances have been made in the design, operation, and optimization of energy communities, practical solutions tailored to residential and citizen-led initiatives remain scarce. In recent years, a growing body of literature has focused on maximizing the value generated within energy communities. Dal Cin et al. [10] proposed multi-objective optimization methodologies that balance economic performance, emissions reduction, and energy efficiency while evaluating different configurations of storage, demand management, and prosumer participation. Koirala et al[11]. highlighted the importance of integrated community energy systems, emphasizing distributed energy resources, flexibility, system integration, and local governance as key elements for collaborative energy infrastructures. Similarly, Ponnaganti et al.[12] demonstrated that demand flexibility and smart control strategies can significantly increase self-sufficiency and facilitate the integration of renewable energy sources.

From a technical design perspective, Soussi et al. [13] reviewed integrated approaches for improving the interoperability and efficiency of rural energy systems through the coordinated optimization of distributed generation and storage. At the same time, advances in digital technologies[14], including IoT platforms[15], energy management systems (EMS) [16,17], and real-time monitoring tools[17], have expanded the capabilities of energy communities. These systems can aggregate distributed resources and operate as virtual power plants, enabling communities to participate more actively in energy markets and provide energy services.

Despite this growing body of literature, the review conducted by Soussi et al. [13] identified more than 114 optimization approaches published between 2020 and 2024.

However, most were developed for technically sophisticated environments and assume levels of expertise, automation, and regulatory flexibility that are rarely available in citizen-led energy communities. Previous studies[18] indicate that approximately one-third of all energy communities fall within the "citizen control" category of Arnstein's ladder of participation, highlighting a high degree of grassroots involvement and local decision-making. These initiatives are typically characterized by installed capacities below 500kWp and are often organized as neighbourhood associations or local cooperatives. This pattern is also evident in Spain, where the National Observatory of Energy Communities reported 837 initiatives by the end of 2025. Among them, 70,8 % had fewer than 50 members, underscoring the prevalence of small-scale, citizen-led models. In most cases (74% in Spain), their primary activity is collective photovoltaic self-consumption, reflecting a practical and accessible entry point into community energy participation. [19]

Within this situation a fundamental question emerges: *which operational and organizational strategies remain available to citizens continue investing in photovoltaic self-consumption systems?* The objective of this study is to identify practical and accessible approaches for citizen-led initiatives capable of improving technical performance, economic resilience, and long-term sustainability while remaining manageable for non-expert users.

To investigate this question, the study focuses on the CERCA Energy Community (Comunidad Energética Renovable de Calatayud), a citizen-led initiative operating across 67 rural municipalities in Aragón, Spain. CERCA currently manages 18 photovoltaic installations with a combined capacity of 845 kWp under Spain's collective self-consumption framework. The analysis is based on real operational and consumption data from one collective self-consumption facility comprising 24 prosumers.

The paper makes three contributions.

The first contribution establishes a counterfactual benchmark to quantify the value created by collective coordination. By comparing the observed community arrangement with a hypothetical scenario of individually operated photovoltaic systems, the analysis isolates the economic and operational benefits that arise from aggregation and coordinated energy use. The results offer evidence to support strategies and policies that can sustain and expand citizen-led photovoltaic initiatives under current market conditions.

The second contribution focuses on reconducting the agreements on shares allocations established energy communities. Although numerous studies have proposed dynamic allocation mechanisms, most have been developed as theoretical optimization exercises or for newly established communities. [20-24] . We also know that this mechanism is not widely adopted by the DSOs in Europe. By contrast, this study evaluates how alternative allocation models can be implemented according to the current legal frameworks in mature communities where investments, governance structures, and participation agreements are already in place. Attention is given to the trade-offs between technical performance, fairness, and practical implementation.

The third contribution examines the economic viability of battery storage integration under current market conditions. Despite its potential to increase self-consumption and flexibility, storage deployment remains limited among citizen energy communities. This study evaluates whether battery investments can generate sufficient value under different allocation schemes and identifies the conditions under which storage becomes economically attractive for community-led initiatives.

The results aim to provide actionable guidance for citizens, community managers, and policymakers seeking to preserve the value of distributed photovoltaic investments under an increasingly challenging economic environment.

## 2. Methodology

### 2.1 Research rationale

The research approach taken in this study operates under a three-step logic.

To begin with, we aimed to quantify the value created by collective coordination. A counterfactual scenario was therefore developed in which each household operated an individual photovoltaic system sized to its annual demand. The resulting technical and economic performance was compared with that of the existing collective installation, allowing the benefits derived from aggregation, load diversity, and coordinated operation to be isolated. This benchmark provides the basis for evaluating the additional value that can be achieved through alternative allocation schemes and storage integration.

Next, we investigated whether the technical parameters, i.e. self-consumption ratio (SCR) or self-sufficiency ratio -SSR-; of the current collective solar plant owned by CERCA could be enhanced by exploring different energy-sharing models. In Spain, the immediate dynamic distribution of energy is not technically applied, creating a significant limitation for optimizing the community's performance. Therefore, we examined four practical allocation models that mimic dynamic functions while still adhering to existing laws. We evaluated these models based on their efficiency, fairness, and possible effects on community governance.

In the final stage, we analyse the integration of a shared battery into the current operational setup to assess how the incorporation of storage affects the technical and financial results for different sharing distribution models. This third step allowed us to quantify how the benefits (or drawbacks) of storage influence users with varying electricity needs, as well as how the selection of the allocation method impacts the investment's value.

All three analytical phases together form a clear framework for not only understanding the community's present performance but also exploring how much more effectively it could operate with different strategies for both operations and investments, creating a practical roadmap to be implemented.

### 2.2 Data and inputs

***Case study:*** The developments presented in this study are framed within the real-world needs and operational constraints faced by energy communities, particularly those that are citizen-led, as opposed to those managed by energy companies. Consequently, the proposed solutions emerge directly from the practical experience of the CERCA Energy Community[25], serving as a foundation for applied and replicable approaches.

***Demand profiles*****:** 24 participating households granted access to their historical electricity consumption records through the DATADIS platform[26]. This provided highly granular (hourly) consumption profiles for two whole years.

***PV shared installation:*** Shared self-consumption photovoltaic installation with a peak power capacity of 116.05 kWp and an inverter rated at 100 kW. The technical specifications of the installation are as follows: a) Photovoltaic modules: Jetion Solar JT550SGH, 550 Wp (211 units); b) Location: 40,42 °N / 3,70 °W; c) Orientation: 60° southwest; d) Tilt angle: 20° and inverter: Growatt MAX 100KTL3-X LV, 100 kW (1 unit). (Figure 1)

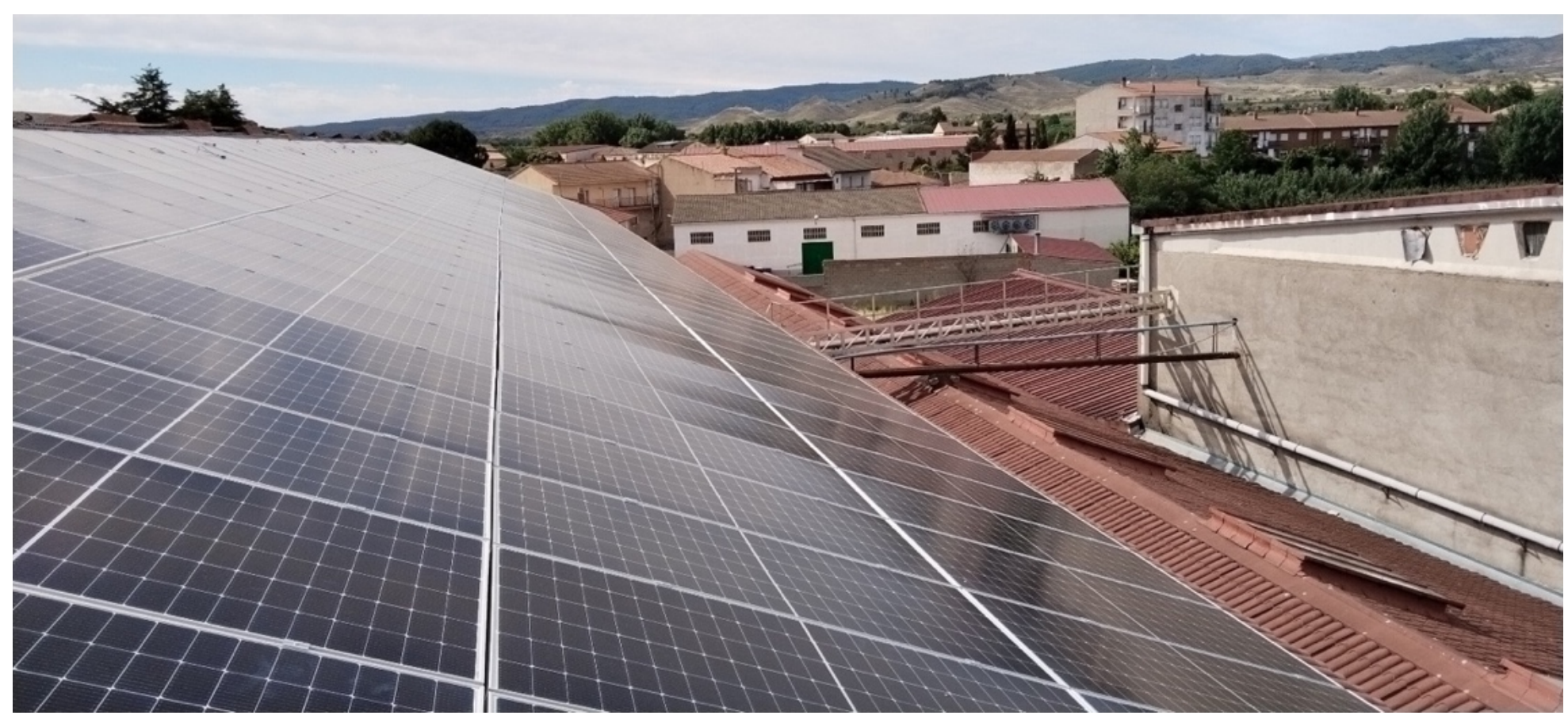

*Figure 1. Image from the installation linked to this study.*

***Energy sharing agreement:*** Table 1 shows the agreement on energy distribution reached by the neighbours, which is just directly proportional to their investment on the facility. The assignment made by local installers is based on a rough estimation regarding the annual demand, without any hourly, daily or seasonal optimization.

***Electricity Prices:*** Electricity prices for both import and export were obtained from the official Red Eléctrica de España (REE) platform (ESIOS), using real hourly market prices.

***Photovoltaic generation:*** It was simulated using hourly data from PVGIS 5 tool for the specific geographical location of the building, assuming the orientation and tilt of current facility. Irradiance data were sourced from the SARAH-3 satellite database.

***PV costs assumptions:*** CAPEX values used in the simulations were not based on generic estimates. Instead, they were derived from real-world price quotes provided by regional PV installers. A linear regression was performed on this dataset to obtain a cost–capacity curve that reflects local market conditions and the economies of scale associated with larger installations. Annual operation and maintenance (O&M) costs were estimated at 12 € $kWp^{-1}yr^{-1}$, accounting for inverter servicing, insurance, and general system upkeep. A Weighted Average Cost of Capital (WACC) of 5% was applied to all investment-related calculations. This discount rate reflects a typical financing structure for cooperative or community-led rural renewable energy projects in Spain, balancing moderate risk with long-term public interest objectives.

| ID profile | Annual demand ($kWh\ yr^{-1}$) | Current assigned percentage (%) | Power assigned (kWp) |
|---|---|---|---|
| **1FB0F** | 3246 | 1.44 | 1.68 |
| **1HK0F** | 20704 | 6.73 | 7.84 |
| **1JCOF** | 2648 | 1.92 | 2.24 |
| **1JSOF** | 2314 | 1.44 | 1.68 |
| **1JYOF** | 3805 | 1.92 | 2.24 |
| **1MLOF** | 40186 | 21.15 | 24.64 |
| **1PAOF** | 1298 | 0.96 | 1.12 |
| **1PHOF** | 54885 | 32.21 | 37.52 |
| **1QDOF** | 3492 | 1.44 | 1.68 |
| **1QV0F** | 1634 | 0.96 | 1.12 |
| **1QXOF** | 2790 | 1.92 | 2.24 |
| **1RYOF** | 7379 | 3.85 | 4.48 |
| **1TGOF** | 2386 | 1.92 | 2.24 |
| **1TJOF** | 3759 | 2.98 | 3.36 |
| **1WW0F** | 2901 | 1.92 | 2.24 |
| **1YT0F** | 4070 | 2.40 | 2.8 |
| **1ZCOF** | 3794 | 1.44 | 1.68 |
| **1ZLOF** | 2096 | 1.92 | 2.24 |
| **1ZM0F** | 4637 | 2.98 | 3.36 |
| **1NG0F** | 1911 | 1.44 | 1.68 |
| **3MR0F** | 2201 | 0.96 | 1.12 |
| **3VCOF** | 2691 | 1.44 | 1.68 |
| **41NG0F** | 2438 | 0.96 | 1.12 |
| **5PJ0F** | 14736 | 3.85 | 4.48 |

*Table 1. Current distribution fixed by the EC with fixed shares based on equity.*

***Technical assumptions:*** Module degradation was assumed to occur linearly at a rate of 0.4% per year, consistent with manufacturer warranties and field data for crystalline silicon technology. The lifetime of the system was set at 20 years, a commonly adopted economic horizon in distributed generation analysis.

***Battery parameters:*** The storage system modelled is a commercial battery (anonymized model) with a maximum charge/discharge power of 107.5 kW, charge and discharge efficiencies of 95.13% and 96.04%, and a usable capacity of 172 kWh (minimum: 43 kWh, maximum: 215 kWh). The initial SOC is set at 78%, which corresponds to 135 kWh of usable capacity, and a lifespan of 7,000 cycles is assumed. Regarding costs for battery use, charging costs are set at €0.126 kWh$^{-1}$ (including efficiency losses and degradation), and discharging at €0.036/kWh.

Table 2 collects source, resolution and main assumptions and uncertainty of data managed.

| Input | Source | Resolution | Assumptions and uncertainty treatment |
|---|---|---|---|
| **Household demand profiles (24 users)** | DATADIS | Hourly | Missing data <2%; linear interpolation applied; inter-annual variability ±4.8–9.3% |
| **PV generation** | PVGIS 5 (SARAH-3 DATABASE) | Hourly | Irradiance variability ±5%; actual location, orientation and tilt used |
| **Import prices** | REE–ESIOS | Hourly | Market volatility captured in sensitivity analysis (−30% to +100%) |
| **Export prices** | REE–ESIOS | Hourly | Wholesale market remuneration; high volatility |
| **PV CAPEX** | Local installer quotations | — | Derived from local quotes using cost–capacity regression; variability ±12% |
| **PV O&M** | Literature + installer data | Annual | Fixed at 12 €/kWp/year |
| **PV Degradation** | Manufacturer warranties + literature | Annual | Linear degradation rate of 0.4%/year |
| **PV Lifetime** | Economic assumption | — | 20-year project horizon |
| **Battery technical parameters** | Manufacturer datasheet | 15 min | Charge/discharge efficiency 95.13–96.04%; usable capacity 172 kwh; lifetime 7,000 cycles |
| **Battery cost** | Market data | — | Sensitivity range 120–500 €/kWh |
| **Discount rate** | Community finance assumptions | — | WACC = 5%, representative of cooperative rural projects |

*Table 2. Details on the data and assumptions used*

## 2.3 Modelling Environment for PV Sizing and Energy Balances

As for the modelling, we utilized OEMOF (Open Energy Modelling Framework), which is an open-source tool based on Python, created for simulating and optimizing intricate energy systems. This tool allows for the creation of mixed linear optimization (MPLI) problems by illustrating system elements, like solar energy generation units, consumption patterns, grid import and export, as well as storage; seen as nodes in a graph framework. In this study, OEMOF played a crucial role in analysing both the energy balance as well as the techno-economic performance under two different scenarios: standalone facilities and a combined facility. It is important to note that OEMOF itself is not a solver; rather, it requires integration with an external optimization solver. For this purpose, we employed Gurobi academic license.

The system was represented as a straightforward pipe labeled "electricity," into which energy from a source (the solar power plant) flows, while the sinks (the users) are directly linked, drawing the necessary energy as much as they want or can take. This pipe is also linked to a different source and sink, which represents the distribution network for the electricity system.

The following assumptions on the modelling and parameter values were fixed.

- The energy system was modelled using the oemof.solph library, assuming perfect foresight over the simulation horizon and no forecasting errors. All simulations were conducted with hourly resolution over a full calendar year (2023/2024), based on representative historical consumption and generation data. The model allowed both grid import and export, with no technical or contractual constraints other than capacity limits.
- The optimization function was designed to minimize the total annual electricity cost for the simulated facilities, which is achieved by evaluating, at each time step, the sum of the costs associated with purchasing electricity from the grid and generating electricity on-site, while subtracting the revenues obtained from selling surplus energy back to the grid.

$$C_{\text{total}}= \sum_t \left[C_{buy,t} + C_{\text{production},t} - I_{\text{sell,t}}\right] \text{ (Eq.1)}$$

Where:

- $C_{total}$ is the total net energy cost over the analysis period. It includes electricity purchase costs from the grid, local production costs, and deducts revenues from energy sold to the grid.
- $t$ is the time index, which is represented in hourly intervals.
- $C_{buy,t}$ is the cost of purchasing electricity from the grid at time $t$. Calculated as:

$$C_{buy,t} = P_{buy,t} \cdot z_{i,t} \text{ (Eq.2)}$$

- $P_{buy}$ t is the market electricity price at time $t$, and $z_{i,t}$ is the amount of energy purchased by user $i$.
- $C_{production,t}$ (LCOE) is representing the cost of local energy production ($C_t$), which accounts for photovoltaic generation capital expenditures (CAPEX), operational expenditures (OPEX), and the discount rate $r$ over a system lifetime of $N$ years.

$$C_{\text{production,t}}= \frac{\sum_{t=1}^{N} \frac{C_t}{(1+r)^t}}{\sum_{t=1}^{N} \frac{E_t}{(1+r)^t}} \text{ (Eq.3)}$$

- $I_{sell,t}$ : Revenue from selling electricity to the grid at time t. Calculated as:

$$I_{sell,t} = P_{sell,t} \cdot w_{i,t} \text{ (Eq.4)}$$

$P_{sell,t}$ is the feed-in tariff or market selling price, and $w_{i,t}$ is the amount of surplus energy exported.

- In scenarios considering PV installation, the investment cost was incorporated as a fixed value amortized linearly over the assumed lifespan of the system, without considering any residual value at the end of the period.

$$Savings = C_{baseline} - C_{optimized} \text{ (Eq.5)}$$

Where:

- $C_{baseline}$: Total electricity cost in the reference scenario (without PV)
- $C_{optimized}$: Total electricity cost under the optimized scenario (with PV)

- No environmental externalities were included.
- The system boundaries were restricted to electrical energy. Thermal uses and other forms of self-generation were not included in the analysis.

### 2.4 Energy-Sharing Models

Distinct methods for sharing allocations were examined for the energy community. Each approach establishes the way in which the photovoltaic (PV) energy production is distributed among those involved, which affects not only technical results (such as SCR and SSR) but also financial aspects (like savings and fairness).

• *Set allocation according to investment (traditional method):* Annual rates remain consistent and are based solely on what each member has invested financially.

• *Set allocation using past dynamic distribution:* Shares are based on the hourly distribution that corresponds to actual usage. Since many regions do not permit real-time dynamic sharing, this approach uses the dynamic hourly distribution that would have taken place in prior years to establish fixed allocations for the group.

• *Set allocation from a past dynamic distribution with surplus sharing*: A significant consideration when using dynamic allocation in shared systems is how to allocate excess energy. This approach utilizes a dynamic allocation profile from a prior year while incorporating a guideline for redistributing any surplus generation among participants.

• *Set allocation based on Marginal Contribution to Profit (MCP):* Shares are calculated based on each member's marginal contribution to the community's savings, determined by running the model again without the demand of that specific member to assess their unique impact.

• *Set allocation based on investment and Marginal Contribution to Profit (MCP):* This mixed strategy merges the small input from everyone towards community savings with their starting funding in the shared solar power facility. While this technique might not suit recently created energy communities, it can act as a temporary fix for long-standing communities wanting to move from entirely investment-focused guidelines without fully switching to a model based on marginal cost pricing.

Dynamic methods: modelling considerations. Energy distribution in the dynamic shared model is distributed according to the following equation:

$$PV_{j,t} = \frac{Load_{j,t}}{\sum_i Load_{i,t}} PV_t \text{ (Eq.6)}$$

Where:

- $PV_{j,t}$ is the amount of photovoltaic (PV) energy allocated to the user $j$ at time $t$. This represents the share of total generated PV energy assigned to user $j$ based on their relative demand.
- $Load_{j,t}$ is the electricity demand of user $j$ at time $t$.
- $Load_{i,t}$, is the total combined demand of all users $i$ at time t. This is used to calculate each user's proportional share of the PV generation.
- $PV_t$ is the total photovoltaic energy generated and available at time $t$ by the shared system.

Two core rules governed:

- When PV generation exceeded community demand, each user covered their full load, and surplus energy was redirected to a "community" sink representing the collective. This energy could be sold or managed for redistribution later.
- When generation fell short of demand, PV output was allocated in proportion to each user's investment share, while remaining energy needs were met via grid import. Export constraints ensured that no more energy was sold to the grid than was generated.

Marginal Contribution to Profit (MCP): modelling considerations. The MCP method assesses the extent of everyone's input to the overall savings achieved by the group. For every user k, the optimization model operates two times: first, including all users and second, omitting user k's load. The additional contribution is then determined by:

$$MCP_k = S_{all} - S_{all-k} \text{ (Eq.7)}$$

Here, $S_{all}$ represents the total community savings with all users included, while $S_{all-k}$ represents the savings without user k. The total community savings ($S_{all}$) are calculated as the difference between the total electricity costs in the baseline scenario (without a photovoltaic system) and the optimized electricity costs when all users participate in the shared energy model. This value reflects the overall economic benefit of the energy community compared to conventional, non-cooperative energy use.

A positive MCP indicates that the user contributes to increasing overall community savings; a negative MCP suggests that their presence reduces collective efficiency, potentially justifying differentiated remuneration or redistribution strategies.

### 2.5 Methodological Framework for the battery

An optimization model based on Mixed-Integer Linear Programming (MILP) to assess optimal energy management within a community energy system featuring a shared storage solution. The objective is to minimize the individual energy cost of each participant, while enforcing proportional limits on storage capacity according to each user's allocation coefficient.

The model employs the Gurobi Optimizer solver, selected for its high performance on large-scale MILP problems and its ability to handle extensive variable sets and constraints. The temporal scope covers a full year (365 days) at 15-minute intervals, allowing for a detailed representation of renewable generation and consumption patterns across daily and seasonal cycles, while ensuring computational feasibility.

The mathematical model includes seven decision variables per user and time step, totaling 245,280 variables per participant. Continuous variables include the energy charged into the battery ($x_{i,t}$) and discharged from it ($y_{i,t}$), both constrained by a maximum energy value M=26.875 kWh. This limit corresponds to the maximum charge and discharge in a 15-minute interval for the battery, which has a nominal capacity of 215 kWh and a maximum charge/discharge rate of 0.5 C (107.5 kW); thus, for a 0.25 h period, the maximum transferable energy is 107.5 kW·0.25 h=26.875 kWh. Additional continuous variables include the energy purchased from (($z_{i,t}$) and sold to ($w_{i,t}$ ) the grid, which have no upper bounds. The state of charge ($SOC_{i,t}$) is another continuous variable, constrained within user-specific limits proportional to their equity coefficients.

Two binary variables per time step —$b_{\text{charge},i,t}$, $b_{\text{discharge},i,t}$ —are introduced to prevent simultaneous charging and discharging, ensuring physically realistic battery operation.

The model incorporates a set of constraints to ensure that the solutions are physically feasible and economically coherent. The fundamental constraint is the energy balance for each user at every time step, which requires that the sum of energy inputs equals the sum of energy outputs. Specifically, the total of photovoltaic (PV) generation, battery discharge adjusted for efficiency, and electricity purchased from the grid must equal the user's total demand, plus the energy used to charge the battery (also adjusted for efficiency) and the energy exported to the grid.

$$Energy\ Inputs\ (sources) = Energy\ Outputs (sinks)$$

$$E_{\text{prod},i,t} + y_{i,t} \cdot \eta_{\text{discharge}} + z_{i,t} = D_{i,t} + \frac{x_{i,t}}{\eta_{\text{charge}}} + w_{i,t} \quad \text{(Eq. 8)}$$

Where:

- $E_{prod,i,t}$ is the PV energy production for user *i* at time *t*
- $y_{i,t}$ is the energy discharged from the battery for user *i* at time *t*
- $\eta_{discharge}$ is the battery discharge efficiency (in this case, 96.04%)
- $z_{i,t}$ is the energy purchased from the grid by user *i* at time *t*
- $D_{i,t}$ is the electrical demand of user *i* at time *t*
- $x_{i,t}$ is the energy charged in the battery for user *i* at time *t*
- $\eta_{charge}$: battery charging efficiency (here, 95.13%)
- $w_{i,t}$: energy sold to the grid by user *i* at time *t*

The state of charge (SOC) of the battery is dynamically linked over time to ensure continuity. At the initial time step:

$$SOC_{i,1} = SOC_{init,i} + x_{i,1} \cdot \eta_{charge} - \frac{y_{i,1}}{\eta_{discharge}} \quad \text{(Eq. 9)}$$

Where:

- $SOC_{i,1}$ is the state of charge for user *i* at the first time period
- $SOC_{init,i}$ is the initial state of charge for user *i*

For all subsequent time steps:

$$SOC_{i,t} = SOC_{i,t-1} + x_{i,t} \cdot \eta_{charge} - \frac{y_{i,t}}{\eta_{discharge}} \quad \text{(Eq. 10)}$$

To prevent physically implausible behaviour, such as charging and discharging simultaneously, binary variables are introduced:

$$x_{i,t} \leq M \cdot b_{\text{charge},i,t}$$
$$y_{i,t} \leq M \cdot b_{\text{discharge},i,t}$$
$$b_{\text{charge},i,t} + b_{\text{discharge},i,t} \leq 1$$

Where:

- *M*: maximum charging/discharging energy (26.875 kWh)
- $b_{charge,i,t}$: binary variable indicating if the battery is charging for user *i* at time *t*
- $b_{discharge,i,t}$: binary variable indicating if the battery is discharging for user *i* at time *t*

Lastly, physical bounds are imposed to ensure feasibility:

- Energy flows (*x,y,z,w*) must be non-negative and $\leq$ maximum capacity.
- SOC must remain within the proportional limits defined for each user.
- All grid interactions must also be non-negative.

The optimization criterion is to minimize the annual energy cost per user. The objective function is as follows:

$$\sum_t \left[P_{\mathrm{buy},t} \cdot z_{i,t} - P_{\mathrm{sell},t} \cdot w_{i,t} + C_{\mathrm{charge}} \cdot x_{i,t} + C_{\mathrm{discharge}} \cdot y_{i,t}\right] \text{ (Eq. 11)}$$

This function accounts for several cost and revenue components:

- $P_{buy,t}$ is the hourly market price of electricity purchased from the grid at time $t$,
- $z_{i,t}$ is the amount of energy user $i$ buys from the grid at time $t$.
- $P_{sell,t}$ is the hourly price received for energy sold back to the grid at time $t$,
- $w_{i,t}$ is the energy that user $i$ sells to the grid at time $t$.
- $C_{charge}$ is the cost per kWh associated with charging the battery, which includes efficiency losses and degradation costs
- $x_{i,t}$ is the energy used to charge the battery by user $i$ at time $t$.
- $C_{discharge}$ is the cost per kWh associated with discharging the battery (also accounting for losses and degradation)
- $y_{i,t}$ is the energy discharged from the battery by user $i$ at time $t$.

The battery storage capacity is shared proportionally among users. Each user optimizes their own consumption and storage strategy independently, within their allocated capacity limits. This structure allows for the evaluation of how different production and consumption profiles influence the individual economic benefits of shared energy storage in a community setting.

For the battery model, the reported savings represent the reduction in electricity costs compared to a scenario with a collective photovoltaic (PV) plant but without storage. These savings are calculated as described in equation 5, where:

- $C_{baseline}$ is the total electricity cost in the scenario with collective PV but without battery storage
- $C_{optimized}$ is the total electricity cost in the scenario with both PV and battery storage

**Economic Feasibility and Sensitivity Analysis**

As a complement to the energy optimization model, an economic evaluation of the community storage system is carried out by calculating the Internal Rate of Return (IRR), along with sensitivity analyses to understand how outcomes change under different market scenarios -uncertainty in prices (±30 to 100%) and battery costs (±50%)-

The economic analysis is based on the annual savings that the community would achieve, according to the model results. These data are used to project cash flows over a 15-year period, which approximately corresponds to the useful life of a lithium-ion battery system in stationary applications. The initial investment is estimated based on a baseline cost of €250/kWh, applied to the system's total usable capacity.

The IRR is calculated using the irr() function from the numpy-financial library, which determines the discount rate that makes the net present value of cash flows equal to zero. In this analysis, the initial cash flow corresponds to the total cost of the storage system (a negative value), followed by positive annual flows equivalent to the energy savings generated. No residual value is included at the end of the period, making this a conservative and cautious profitability estimate.

# 3. Results and discussion

## 3.1 Individual self-consumption approach

The purpose of this comparison is not to evaluate future decisions, but to establish a reference scenario against which the value generated by collective coordination can be quantified. The individual self-consumption case therefore serves as a counterfactual benchmark representing a fragmented operation of the same users and demand profiles.

In this first simulation, the PV installation for each user was sized according to their individual annual electricity demand. The system design was based on the technical specifications of the PV modules used in the real project, ensuring both realism and consistency. Each system's capacity was adjusted in discrete steps corresponding to whole units of these modules, resulting in technically feasible system sizes. The performance and financial results for these individually sized PV systems are summarised in Table 3.

| ID Profile | Annual Demand (kWh $yr^{-1}$) | PV Size (kWp) | Costs(€ kWp ) | Investment (€) | Cost of imported energy (€ $yr^{-1}$) | Revenue from exported energy (€ $yr^{-1}$) | Energy Imported (KWh $yr^{-1}$) | Energy Exported (Wh $yr^{-1}$) | Energy Produced (kWh $yr^{-1}$) | LCOE (€ $kWh^{-1}$) | Energy costs without PV facility (€ $yr^{-1}$) | Annual Energy Costs with PV facility (€ $yr^{-1}$) | Savings (€ $yr^{-1}$) | Savings (% $yr^{-1}$) | SSR(%) | SCR(%) |
|---|---|---|---|---|---|---|---|---|---|---|---|---|---|---|---|---|
| 1FB0F | 3246 | 2 | 1743 | 3486 | 323.1 | 161.2 | 2153 | 2208 | 3301 | 0.0528 | 467.3 | 336.2 | 131.1 | 28% | 33.66 | 33.11 |
| 1HK0F | 20704 | 13 | 1085 | 14105 | 2006.3 | 924.0 | 12734 | 13492 | 21462 | 0.0329 | 3146.2 | 1787.5 | 1358.8 | 43% | 38.49 | 37.12 |
| 1JC0F | 2648 | 2 | 1743 | 3486 | 249.7 | 162.9 | 1610 | 2263 | 3301 | 0.0528 | 390.7 | 261 | 129.7 | 33% | 39.20 | 31.45 |
| 1JS0F | 2314 | 1.5 | 1764 | 2646 | 243.8 | 126.6 | 1582 | 1745 | 2477 | 0.0534 | 345.8 | 249.6 | 96.3 | 28% | 31.63 | 29.53 |
| 1JY0F | 3805 | 2.5 | 1721 | 4304 | 344.6 | 183.4 | 2217 | 2538 | 4126 | 0.0521 | 562.3 | 376.2 | 186.1 | 33% | 41.74 | 38.48 |
| 1ML0F | 40186 | 24.5 | 1028 | 25174 | 3513.4 | 1299.5 | 20363 | 20613 | 40436 | 0.0311 | 6726.8 | 3472.4 | 3254.4 | 48% | 49.33 | 49.02 |
| 1PA0F | 1298 | 1 | 1786 | 1786 | 116.3 | 77.7 | 761 | 1114 | 1651 | 0.0541 | 191.6 | 127.9 | 63.7 | 33% | 41.37 | 32.53 |
| 1PH0F | 54885 | 33.5 | 983 | 32914 | 4155.8 | 1495.2 | 25058 | 25467 | 55294 | 0.0298 | 8940.8 | 4306.1 | 4634.7 | 52% | 54.34 | 53.94 |
| 1QD0F | 3492 | 2.5 | 1721 | 4304 | 310.1 | 187.3 | 1954 | 2589 | 4127 | 0.0521 | 521.0 | 337.9 | 183.1 | 35% | 44.04 | 37.26 |
| 1QV0F | 1634 | 1 | 1786 | 1786 | 144.6 | 69.7 | 931 | 948 | 1651 | 0.0541 | 239.6 | 164.7 | 74.9 | 31% | 43.01 | 42.58 |
| 1QX0F | 2790 | 2 | 1743 | 3486 | 256.3 | 151.4 | 1646 | 2157 | 3301 | 0.0528 | 419.1 | 279.20 | 139.9 | 33% | 41.01 | 34.66 |
| 1RY0F | 7379 | 4.5 | 1336 | 6011 | 750.2 | 311.1 | 4689 | 4444 | 7134 | 0.0421 | 1719.8 | 1251.6 | 468.2 | 27% | 39.76 | 38.49 |
| 1TG0F | 2386 | 1.5 | 1764 | 2646 | 240.5 | 110.7 | 1462 | 1552 | 2476 | 0.0534 | 364.6 | 262.0 | 102.5 | 28% | 38.73 | 37.32 |
| 1TJ0F | 3759 | 2.5 | 1721 | 4304 | 356 | 194.7 | 2327 | 2695 | 4127 | 0.0521 | 547.5 | 376.4 | 171.1 | 31% | 38.08 | 34.69 |
| 1WW0F | 2901 | 2 | 1743 | 3486 | 262.6 | 157.4 | 1766 | 2166 | 3301 | 0.0528 | 417.07 | 279.4 | 137.7 | 33% | 39.13 | 34.38 |

| 1YT0F | 4070 | 2.5 | 1721 | 4304 | 374.0 | 182.3 | 2461 | 2518 | 4127 | 0.0521 | 590.0 | 406.8 | 183.2 | 31% | 39.52 | 38.97 |
|---|---|---|---|---|---|---|---|---|---|---|---|---|---|---|---|---|
| 1ZC0F | 3794 | 2.5 | 1721 | 4304 | 321.3 | 173.2 | 2079 | 2412 | 4127 | 0.0521 | 558.5 | 363.2 | 195.3 | 35% | 45.19 | 41.54 |
| 1ZL0F | 2096 | 1.5 | 1764 | 2646 | 216.3 | 129.0 | 1402 | 1782 | 2476 | 0.0534 | 308.6 | 219.6 | 89.0 | 29% | 33.1 | 28.01 |
| 1ZM0F | 4637 | 3 | 1700 | 5100 | 448.2 | 228.6 | 2860 | 3176 | 4953 | 0.0515 | 686.2 | 475.0 | 211.2 | 31% | 38.3 | 35.85 |
| 1NG0F | 1911 | 1.5 | 1764 | 2646 | 172.9 | 115.8 | 1101 | 1667 | 2477 | 0.0534 | 288.7 | 189.4 | 99.3 | 34% | 42.35 | 32.68 |
| 3MR0F | 2201 | 1.5 | 1764 | 2646 | 187.1 | 104.8 | 1185 | 1461 | 2477 | 0.0534 | 329.3 | 214.5 | 114.7 | 35% | 46.15 | 41.00 |
| 3VC0F | 2691 | 2.0 | 1743 | 3486 | 230.9 | 160.3 | 1573 | 2184 | 3302 | 0.0528 | 380.6 | 244.8 | 135.8 | 36% | 41.51 | 33.84 |
| 41NG0F | 2438 | 1.5 | 1764 | 2646 | 292.9 | 134.1 | 1869 | 1908 | 2477 | 0.0534 | 373.0 | 291.1 | 81.9 | 22% | 23.34 | 22.94 |
| 5PJ0F | 14736 | 9.0 | 1143 | 10286 | 1555.2 | 721.5 | 10373 | 10501 | 14864 | 0.0346 | 2182.5 | 1347.9 | 834.6 | 38% | 29.57 | 29.3 |
| TOTAL | 192002 | 121 | - | 151986 | 17071.9 | 7561.6 | 106156 | 113600 | 199446 | - | 30697.7 | 17620.4 | 13077.2 | 43% | - | - |

*Table 3. Design for individual self-consumption.*

0

1

Capital expenditures (CAPEX) were calculated for each user according to the size of their respective PV system sized by OEMOF. These values reflect realistic market costs, which include not only solar modules but also inverters, mounting structures, other balance-of-system components and installation costs. As commonly observed, economies of scale significantly reduced the investment per kilowatt-peak installed for larger systems. Specifically, small systems presented investment values near €1,742.90 $kWp^{-1}$, while larger systems reached levels as low as €982.50 $kWp^{-1}$. As a result, total investment costs ranged from approximately €1,785 for the smallest systems to over €33,000 for the largest.

Energy production for each system was estimated using high-resolution solar irradiation data and typical performance parameters. Annual PV generation ranged from 1,651 kWh to 55,293 kWh per user. Self-consumption and export shares were computed based on the household’s load profile and PV output. The resulting energy balances enabled the calculation of energy imported from the grid, exported to it, and self-consumed.

The Levelized Cost of Electricity (*LCOE*), the average cost per kilowatt-hour of solar energy generated over the system’s lifespan, was also computed for each user. These values provide a comparable measure of long-term cost efficiency. As expected, LCOE values were lower for larger systems (e.g., €0.0298 $kWh^{-1}$ for 33.5 kWp) and higher for small-scale setups (e.g., €0.0541 $kWh^{-1}$ for 1.0 -1.5 kWp), reflecting the influence of scale on amortized investment.

Based on these energy flows and cost assumptions, annual electricity costs were estimated for each user under the self-consumption scenario. This included three main components: the cost of electricity still imported from the grid, the revenue received for exported surplus energy, and the cost of self-consumed PV energy (calculated using the user-specific *LCOE* and the amount of energy retained for internal use). These figures were then compared to the total annual energy cost in the reference scenario, in which users purchase all their electricity from the grid at retail price.

Overall, the results demonstrate substantial savings for nearly all users. Annual savings ranged from €63 to €4,634, depending on system size, consumption pattern, and alignment with solar production. In percentage terms, most users reduced their energy costs by approximately 28 % to 35 %. The largest relative savings, up to 52 %, were achieved by users with large and well-utilized systems.

Attending to the energy usage, it is observed a general alignment between *SSR* and *SCR* across most users, although there are some notable discrepancies. The highest SSR and SCR values are found for industrial consumers whose demand is concentrated during working hours, when PV generation is also at its peak. In contrast, users with consumption patterns that are less aligned with solar production achieve noticeably lower values. In addition, some participants show a marked gap between SSR and SCR, suggesting that although a significant share of their demand is covered by the PV system, they are not always able to directly consume all the energy allocated to them.

### 3.2 Shared self-consumption approach

#### 3.2.1 How much it worth it?

Continuing from the baseline analysis of stand-alone systems, we next simulated a collective scenario in which all households pool their demand and are supplied by a single PV plant. Every technical parameter other than scale as module efficiency, orientation, hourly load traces, was kept identical to the individual case; the optimisation routine was allowed to size the shared array freely. The resulting solution, presented in Table 4, calls for a 110.5 kWp installation, about nine per cent smaller than the total 121 kWp that would be required if each dwelling pursued self-consumption on its own. Because bulk procurement reduces the specific investment to €750 $kWp^{-1}$ and the balance-of-system is consolidated into one site, the collective plant demands an upfront outlay of €82,875 , almost forty-five per cent below the sum of individual investments. That figure is not merely theoretical: the real project in Calatayud eventually commissioned 211 standard 550 W modules (116.05 kWp) with a net cost of €88,039, a number that sits between the modelled optimum (201 units) and the aggregate of stand-alone layouts (242 units). The comparison reveals the first tangible advantage of the energy community shared self-consumption approach: fewer modules and therefore less roof/land area, cabling and inverter capacity are needed to serve the same annual demand.

Operationally, the shared plant generates 176.9 MWh $yr^{-1}$, covering ninety-two per cent of the community's total consumption. Load diversity softens instantaneous peaks, so only 95 MWh must still be imported from the grid, while 90 MWh are exported when midday production exceeds aggregate demand. The higher simultaneity between generation and consumption raises on-site self-consumption to 49,18%, a clear improvement over the 36,2% average obtained under individual operation. Thanks to both scale and improved utilisation, the community achieves a levelized cost of electricity of €0.023 $kWh^{-1}$, roughly one-third lower than the best household system and less than half the mean LCOE of the stand-alone portfolio.

| | |
|---|---|
| **Anual Demand (kWh $yr^{-1}$)** | 192002 |
| **PV Size (kWp)** | 110.5 |
| **Costs(€ kWp )** | 750.0 |
| **Investment (€)** | 82875,0 |
| **Cost of imported energy (€ $yr^{-1}$)** | 15391.8 |
| **Revenue from exported energy (€ $yr^{-1}$)** | 5794.7 |
| **Energy Imported (kWh $yr^{-1}$)** | 95247 |
| **Energy Exported (Wh $yr^{-1}$ )** | 90230 |
| **Energy Produced (kWh $yr^{-1}$)** | 176993 |
| **LCOE (€ $kWh^{-1}$)** | 0.0234 |
| **Payback** | 4.9 |

*Table 4. Optimized design for shared self-consumption.*

The financial benefits of the collective configuration are clear. Without PV, the community would spend around €30,700 per year on electricity. With the shared installation, annual costs fall to €13,741, resulting in savings of almost €17,000 per year.

Based on the modelled investment cost of €82,875, this corresponds to a simple payback period of 4.9 years. Using the actual net investment of the commissioned plant (€88,039), the payback period remains close, at 5.2 years. These values reinforce the economic attractiveness of collective self-consumption even under current electricity market conditions. Savings are around 55% of the variable electricity bill and about 20% higher than the combined savings achieved by the 24 individual systems (€16,956 versus €13,077).

Part of this advantage comes from the diversity of consumption profiles. When users share generation, energy that would otherwise be exported to the grid can often be consumed by another member of the community. As a result, more locally generated electricity is used on-site and less value is lost through low export prices.

It should also be noted that these calculations only consider energy-related costs. Contracted power charges, metering fees and taxes were not included. In practice, community members may achieve additional savings by reducing contracted power levels and benefiting from the regulatory framework for collective self-consumption. Actual economic benefits are therefore likely to be somewhat higher than those reported here.

Beyond the financial results, collective self-consumption also offers practical advantages. A single grid connection simplifies administration and permitting, maintenance can be managed collectively, and the installation retains flexibility to accommodate future members or new loads such as electric vehicle charging.

Most importantly, these results show that a significant share of the value generated by the community comes from coordination rather than from photovoltaic generation alone. When users share a common installation, differences in their consumption patterns allow locally generated electricity to be used more efficiently. Energy that would otherwise be exported to the grid by one participant can often be consumed by another, increasing self-consumption and reducing overall costs.

Although exploring this effect in detail is beyond the scope of this paper, we propose to define the load-profile complementarity or complementarity index to understand benefits to share consumption that could be obtained by users. In simple terms, this indicator, defined in equation 12, would describe how well the demand patterns of different users fit together. It is calculated as the difference between the PV energy consumed under aggregated operation and the sum of PV energy self-consumed by each user operating independently, normalized by the total PV generation. Higher values would indicate greater complementarity among demand profiles and a stronger potential for collective self-consumption benefits.

$$CI = \frac{\sum_t [min(\sum_i D_{i,t}, PV_t) - min(\sum_i D_{i,t}, PV_{i,t})]}{\sum_t [PV_t]} \text{ (eq.12)}$$

Where:

- $D_{i,t}$ = electricity demand of user $i$ at time $t$.
- $PV_{i,t}$ = photovoltaic generation allocated to user $i$ in the individual scenario.
- $PV_t$ = total photovoltaic generation available to the community at time $t$.
- $\min(D_{i,t}, PV_{i,t})$ = PV energy self-consumed by user $i$ in the individual scenario.
- $\min(\sum_i D_{i,t}, PV_t)$ = PV energy self-consumed under collective operation.

The greater this complementarity, the greater the opportunity to convert surplus generation into useful local consumption. This underlying effect helps explain why the collective configuration outperforms the equivalent set of individual installations.

Having quantified the coordination value, the following sections investigate whether additional improvements can still be achieved once the collective asset is already in operation through alternative allocation schemes and storage integration.

### 3.2.2 Benefits coming from diverse allocation shares.

Table 1 showed the allocation share members of energy community studied defined based on the equity investment and. The share is applied uniformly every single hour of the year. Results from such allocation are those commented in the previous section. The next question is to what extent the energy distribution model based solely on equity can be considered optimal.

#### 3.2.2.1 Set allocation using past dynamic distribution

First, it should be acknowledged that dynamic energy sharing, although recognized as the optimal solution by several authors, cannot currently be implemented in many countries, including Spain. Nevertheless, while Spanish regulation requires fixed allocation coefficients, these coefficients may vary across the 8,760 hours of the year.

Accordingly, our first proposal is to define allocation shares based on the hourly consumption patterns observed in previous years. Due to limitations in the DATADIS platform, only the two most recent years of consumption data were available. Analysis of these data revealed an interannual variability in individual consumption ranging from 4.5% to 9.3%. An improved approach would be to construct a representative consumption profile using all available historical data, analogous to the Typical Meteorological Year commonly used in photovoltaic energy modelling.

To explore this, we conducted an optimization using the oemof.solph library, which allows us to model hour-by-hour allocations based on each participant's actual consumption profile.

The objective of this optimization was to minimize costs within the community, meaning that electricity generated by the shared photovoltaic system is consumed locally as much as possible, minimizing both exports to the grid and energy imports. The model respects technical constraints, such as hourly demand data for each user, and ensures that at each moment, the available generation is allocated only among those users who are actively consuming. This pseudo-dynamic allocation scheme offers a more realistic and potentially more efficient alternative to the static, fixed-share approach typically used in collective self-consumption schemes.

Table 5 presents a detailed snapshot of how electricity from the designed shared photovoltaic (PV) installation presented above is distributed among the 24 members of a rural energy community. Instead of showing absolute energy values, the table displays the percentage of available generation allocated to each user during selected hourly intervals, based on the distribution logic embedded in the optimization model.

155

| | **1FB0F** | **1HK0F** | **1JCOF** | **1JSOF** | **1JYOF** | **1MLOF** | **1PAOF** | **1PHOF** | **1QDOF** | **1QV0F** | **1QXOF** | **1RYOF** | **1TGOF** | **1TJOF** | **1WW0F** | **1YT0F** | **1ZCOF** | **1ZLOF** | **1ZM0F** | **1NG0F** | **3MR0F** | **3VCOF** | **41NG0F** | **5PJ0F** | Total |
|---|---|---|---|---|---|---|---|---|---|---|---|---|---|---|---|---|---|---|---|---|---|---|---|---|---|
| Annual Demand (kWh yr⁻¹) | 3245,8 | 20704 | 2648,3 | 2314,1 | 3805,4 | 40186 | 1298,1 | 54885 | 3491,7 | 1634,1 | 2790,1 | 7379 | 2386,1 | 3759,2 | 2901,3 | 4070,1 | 3794 | 2096 | 4637 | 1911 | 2201 | 2691 | 2438 | 14736 | |
| Energy allocation based on previous demand shown in % for different selected days and hours along the year | | | | | | | | | | | | | | | | | | | | | | | | | |
| **2023-01-01 15:00:00** | 14,50 | 25,30 | 2,88 | 1,84 | 8,14 | 5,45 | 0,95 | 5,45 | 3,73 | 0,15 | 1,66 | 0,00 | 0,39 | 2,69 | 0,92 | 7,14 | 3,62 | 0,54 | 3,20 | 1,14 | 0,60 | 3,69 | 0,29 | 5,73 | 100,0 |
| **2023-01-01 16:00:00** | 17,01 | 25,40 | 2,17 | 0,35 | 4,31 | 0,00 | 0,95 | 10,81 | 1,47 | 0,16 | 1,65 | 0,00 | 0,43 | 4,56 | 6,28 | 4,53 | 1,61 | 0,50 | 2,29 | 1,14 | 0,59 | 2,96 | 1,07 | 9,78 | 100,0 |
| **2023-02-25 10:00:00** | 0,22 | 7,31 | 0,24 | 0,00 | 0,94 | 1,28 | 0,14 | 2,57 | 0,40 | 0,36 | 0,38 | 0,00 | 0,08 | 0,26 | 0,17 | 0,45 | 0,34 | 0,09 | 0,55 | 0,28 | 0,98 | 0,07 | 1,45 | 4,37 | 22,93 |
| **2023-03-02 11:00:00** | 0,15 | 6,91 | 0,18 | 0,00 | 0,23 | 22,67 | 0,13 | 29,47 | 0,21 | 0,21 | 0,46 | 1,80 | 0,14 | 0,24 | 0,23 | 0,34 | 0,35 | 0,11 | 0,53 | 0,46 | 0,24 | 0,07 | 0,74 | 4,37 | 70,27 |
| **2023-03-09 07:00:00** | 1,44 | 18,56 | 1,40 | 0,00 | 1,56 | 20,77 | 0,60 | 31,16 | 0,86 | 0,69 | 1,52 | 1,51 | 1,60 | 1,29 | 0,83 | 2,11 | 1,67 | 0,83 | 1,51 | 1,00 | 0,55 | 0,53 | 0,00 | 8,01 | 100,0 |
| **2023-03-19 18:00:00** | 1,42 | 11,68 | 1,89 | 0,00 | 2,29 | 7,82 | 0,90 | 15,65 | 11,14 | 1,59 | 2,32 | 0,00 | 2,01 | 2,30 | 6,86 | 2,07 | 2,00 | 1,59 | 2,28 | 1,51 | 1,33 | 0,41 | 17,54 | 3,39 | 100,0 |
| **2023-06-08 09:00:00** | 1,05 | 6,09 | 1,39 | 4,04 | 1,57 | 18,07 | 0,96 | 45,17 | 1,40 | 0,93 | 1,27 | 4,87 | 0,73 | 1,29 | 0,75 | 2,41 | 1,78 | 0,82 | 1,76 | 1,14 | 1,22 | 0,54 | 0,00 | 0,75 | 100,0 |
| **2023-07-06 16:00:00** | 4,13 | 2,86 | 0,76 | 0,74 | 0,83 | 7,64 | 0,54 | 25,45 | 0,69 | 0,55 | 0,81 | 0,85 | 0,32 | 3,02 | 0,70 | 1,13 | 0,91 | 0,75 | 0,97 | 0,64 | 0,63 | 0,14 | 0,09 | 1,14 | 56,29 |
| **2023-08-23 06:00:00** | 2,31 | 4,41 | 3,58 | 0,00 | 2,02 | 0,00 | 3,22 | 28,90 | 2,50 | 1,37 | 4,55 | 0,00 | 1,45 | 3,35 | 1,85 | 6,73 | 3,84 | 3,27 | 4,10 | 2,85 | 1,21 | 0,84 | 0,00 | 17,63 | 100,0 |
| **2023-08-23 18:00:00** | 1,03 | 1,75 | 0,99 | 0,00 | 1,27 | 56,42 | 0,85 | 22,57 | 0,73 | 0,37 | 1,58 | 0,00 | 0,35 | 0,71 | 0,50 | 0,97 | 0,79 | 2,08 | 1,64 | 0,78 | 0,55 | 0,35 | 0,00 | 3,71 | 100,0 |
| **2023-10-19 16:00:00** | 1,34 | 5,70 | 1,25 | 2,64 | 2,06 | 18,83 | 0,34 | 45,20 | 2,44 | 0,50 | 0,98 | 0,00 | 1,44 | 2,93 | 3,19 | 1,24 | 1,82 | 1,50 | 1,52 | 0,77 | 0,41 | 3,01 | 0,18 | 0,71 | 100,0 |
| Shares based on equity investment | 1,44 | 6,73 | 1,92 | 1,44 | 1,92 | 21,15 | 0,96 | 32,21 | 1,44 | 0,96 | 1,92 | 3,85 | 1,92 | 2,88 | 1,92 | 2,40 | 1,44 | 1,92 | 2,88 | 1,44 | 0,96 | 1,44 | 0,96 | 3,85 | 100,0 |

| Pure Dynamic model | 1,21 | 6,52 | 0,97 | 0,67 | 1,38 | 14,00 | 0,54 | 21,49 | 1,28 | 0,65 | 1,07 | 1,09 | 0,87 | 1,50 | 1,00 | 1,62 | 1,51 | 0,90 | 1,67 | 0,73 | 0,83 | 1,03 | 0,76 | 3,73 | 67,02 |
|---|---|---|---|---|---|---|---|---|---|---|---|---|---|---|---|---|---|---|---|---|---|---|---|---|---|
| Blended dynamic | 1,80 | 9,73 | 1,45 | 1,01 | 2,06 | 20,88 | 0,80 | 32,07 | 1,91 | 0,98 | 1,60 | 1,62 | 1,29 | 2,24 | 1,49 | 2,41 | 2,26 | 1,34 | 2,49 | 1,09 | 1,23 | 1,54 | 1,13 | 5,57 | 100,0 |
| Difference | 0,36 | 3,00 | - 0,47 | - 0,44 | 0,14 | -0,27 | - 0,16 | -0,14 | 0,46 | 0,01 | - 0,32 | - 2,22 | - 0,63 | - 0,64 | - 0,43 | 0,01 | 0,81 | - 0,58 | - 0,40 | - 0,35 | 0,27 | 0,10 | 0,17 | 1,72 | 0,00 |

*Table 5. Optimized shares are based on historical consumption patterns. Columns represent individual users, while rows labelled with dates indicate the hourly allocation shares assigned to each user under the dynamic model based on past patterns. Fixed shares based on equity investment represent the current approach adopted in the energy community. The 'pure dynamic' model refers to the average annual share allocated to each user based solely on real-time consumption. The 'blended dynamic' model applies surplus redistribution according to equity-based shares layered over the dynamic allocation. Difference is the difference between the fixed shares model and the blended dynamic*

| | **1FB0F** | **1HK0F** | **1JCOF** | **1JSOF** | **1JYOF** | **1MLOF** | **1PAOF** | **1PHOF** | **1QDOF** | **1QV0F** | **1QXOF** | **1RYOF** | **1TGOF** | **1TJOF** | **1WW0F** | **1YT0F** | **1ZCOF** | **1ZLOF** | **1ZM0F** | **1NG0F** | **3MR0F** | **3VCOF** | **41NG0F** | **5PJ0F** | Total |
|---|---|---|---|---|---|---|---|---|---|---|---|---|---|---|---|---|---|---|---|---|---|---|---|---|---|
| Shares based on equity investment | 1,44 | 6,73 | 1,92 | 1,44 | 1,92 | 21,15 | 0,96 | 32,21 | 1,44 | 0,96 | 1,92 | 3,85 | 1,92 | 2,88 | 1,92 | 2,40 | 1,44 | 1,92 | 2,88 | 1,44 | 0,96 | 1,44 | 0,96 | 3,85 | 100 |
| Blended dynamic | 1,80 | 9,73 | 1,45 | 1,01 | 2,06 | 20,88 | 0,80 | 32,07 | 1,91 | 0,98 | 1,60 | 1,62 | 1,29 | 2,24 | 1,49 | 2,41 | 2,26 | 1,34 | 2,49 | 1,09 | 1,23 | 1,54 | 1,13 | 5,57 | 100 |
| MC model . Share (%) | 1,58 | 9,46 | 1,22 | 1,22 | 1,74 | 23,93 | 0,47 | 32,09 | 1,76 | 0,85 | 1,28 | 3,51 | 1,18 | 1,67 | 1,3 | 2,03 | 1,82 | 0,83 | 2,04 | 0,86 | 0,99 | 1,4 | 0,88 | 5,87 | 100 |
| Mixed Equity + MC | 1,51 | 8,1 | 1,57 | 1,33 | 1,83 | 22,55 | 0,72 | 32,15 | 1,6 | 0,91 | 1,6 | 3,68 | 1,55 | 2,28 | 1,61 | 2,22 | 1,63 | 1,38 | 2,46 | 1,15 | 0,97 | 1,42 | 0,92 | 4,86 | 100 |

*Table 6. Optimized shares for the different allocation models*

Out of the 8,760 total hours, the model identified 4,504 hours with no PV generation, leaving 4,256 hours during which energy could be allocated.

The selected time steps in the table illustrate how this dynamic model responds to varying conditions. For example, on March 2nd at 11:00, under high solar output and moderate demand, user 1PHOF receives nearly 29.5% of the energy and, 1MLOF a 22,67%, whereas others, such as 1ZLOF, receive considerably less. On August 23rd at 18:00, despite high demand relative to limited production, the load profile of 1MLOF enables it to absorb up to 56.4% of the available generation.

One notable observation is that even in time steps where 100% of the generation is consumed, not all 24 users are simultaneously drawing their full static assigned share. This reflects the practical reality of asynchronous consumption: some users regularly consume less than their proportional allocation, allowing others to use a greater share of the generation in real time.

The pure dynamic model significantly improves the self-consumption rate, 67.02% of generated energy is used locally, compared to only 50.52% under fixed coefficients. In other words, dynamic allocation boosts self-consumption by nearly 17 percentage points. This improvement is not merely a technical efficiency; it has clear economic significance. In electricity markets where the spread between the retail price of electricity and the remuneration for exported surplus is considerable, increasing self-consumption directly enhances economic returns. Every kilowatt-hour consumed on-site avoids paying the full retail tariff, while surplus energy exported to the grid is typically compensated at much lower rates, often based on fluctuating wholesale prices or capped feed-in tariffs.

Assuming a spread ranging from €0.06 $kWh^{-1}$ to €0.15 $kWh^{-1}$, this 17% increase in self-consumption would translate into additional annual savings of approximately €1,800 to €4,500, depending on market conditions. Thus, beyond technical optimization, dynamic allocation offers a robust financial advantage by capturing more of the retail value of the electricity produced and consumed within the community.

This analysis highlights how more flexible and responsive allocation mechanisms can materially improve the financial performance of collective self-consumption systems, particularly under current regulatory and market conditions with asymmetric buy-sell prices.

#### 3.2.2.2 Set allocation from a past dynamic distribution with surplus sharing

However, this approach leaves 32.98% of annual PV generation unallocated due to unused surpluses. Selling this energy externally acting as an energy retailer could be an option but would introduce legal and operational complexity not considered in this study. Instead, we assume a simple internal solution: redistributing the unused surplus proportionally to each participant's investment share, as shown in the third row "Blended Dynamic" model at the Table 5. This model combines the dynamic allocation based on historical demand patterns with a proportional redistribution of the remaining surplus according to ownership shares. For newly established energy communities, this surplus could instead be allocated directly following the same proportions derived from the dynamic allocation model, as investment decisions can be defined before the sharing agreement enters into force.

This hybrid strategy achieves full utilization of the generated energy while preserving equity among participants. The differences between the current model based on equity and equally distributed along the year and the blended scenarios are revealing. Users such as 1PHOF and 1MLOF show only marginal gains or losses, suggesting their consumption closely matches their fixed share. In contrast, users like 1TJOF, 1RYOF, and 5PJOF experience sharp differences, up to 2.22 percentage points below their static share, indicating they consistently under-consume relative to their assigned portion.

These patterns have real design implications. A purely fixed allocation misaligns with actual consumption behaviour, while a dynamic model, though more efficient reaching a SCR of 67,02%, fails to fully utilize available energy. The blended solution appears optimal but also reveals intra-community asymmetries: some members benefit more from flexibility, while others lose potential access to surplus energy. Nevertheless, its implementation in existing energy communities should be relatively straightforward, as the resulting annual allocation shares remain close to those originally agreed under equity-based schemes.

#### 3.2.2.3 Set allocation based on Marginal Contribution to Profit (MC)

Transitioning from a community energy management model that prioritizes equity through fixed distribution shares based on investment to one that maximizes self-consumption and minimizes costs, raises the question: what would happen if energy distribution were instead based on marginal contribution? This metric captures the added value each participant brings to the community, not only in financial terms but also in how their consumption patterns enhance the overall efficiency of the system. The model preserves the initial investment commitments but redefines energy allocation shares to reflect optimized performance and fairness. As a result, the distribution differs substantially from the current equity-based arrangement.

This added value may come in two forms:

- Cost-effectiveness: A participant may help reduce overall system costs by enabling economies of scale or by contributing capital at key stages.
- Profile complementarity: A user's consumption pattern may align with gaps in other members' demand, thereby boosting the community's overall self-consumption rate.

To explore this idea further, we proposed the evaluation of the marginal contribution (MC) of each participant as a tool to enhance both fairness and efficiency in energy-sharing models within renewable energy communities. The marginal contribution is defined in equation 7 and results are shown in Table 6.

This method is implemented through a set of fixed allocation shares that remain constant throughout the year. Under this allocation scheme, the community reaches a SCR of 60.1% and a SSR of 55.4%. Compared with the current equity-based allocation, the SCR increases by 10.9 percentage points, highlighting the value of considering each participant's contribution to collective performance rather than relying solely on investment shares. Although the MC approach does not achieve the performance of the dynamic allocation models explored above, it provides a substantial improvement while also remaining compatible with the fixed-share requirements of current regulation, and it could be easily to calculate for lay citizens.

From a technical perspective, the MC model better aligns incentives with system performance and can therefore be considered a fair allocation mechanism from an operational standpoint. However, its implementation may be socially challenging in established energy communities. Participants who financed a significant share of the installation may perceive a substantial reduction in their allocation rights despite having assumed a larger financial risk during the project's development. Consequently, although the MC approach improves efficiency, it may create governance tensions if introduced abruptly.

#### 3.2.2.4 Set allocation based on investment and Marginal Contribution to Profit (MC)

This hybrid model was developed to address the main limitation of the previous approaches. While the equity-based model recognizes the financial commitment made by participants, it ignores their actual contribution to the efficient use of community energy. Conversely, the MC model rewards operational value creation but may disregard the investments that enabled the project in the first place. The hybrid approach combines both perspectives by allocating 50% of the shares according to ownership and 50% according to marginal contribution. The objective is not to maximize technical performance alone, but to provide a practical transition pathway for existing energy communities seeking to improve efficiency without undermining previously established investment agreements
.

The hybrid model provides a compromise between efficiency and stability. As shown in Figure 2 and Table 6, the resulting allocations remain relatively close to the current equity-based arrangement while correcting some of its most significant mismatches.
For example, 1HK0F increases from 6.73% in the current scheme to 8.10% in the hybrid model, capturing part of its higher operational contribution without reaching the 9.46% assigned under the pure MC approach. Similarly, 5PJ0F increases from 3.85% to 4.86%, while 1JC0F decreases only moderately from 1.92% to 1.57%, avoiding the sharper reduction observed under the pure MC allocation.

Unlike the pure MC model, the hybrid approach reduces the likelihood of resistance from participants whose allocation would otherwise decrease substantially. At the same time, it captures a significant portion of the efficiency gains associated with performance-based allocation. For mature energy communities, where governance stability and member acceptance are often as important as technical optimization, this approach may therefore represent the most realistic pathway for revising allocation agreements.

### 3.3 Battery storage integration

Battery storage was analysed as a complementary strategy to increase the value of locally generated PV energy within the community. As the energy community currently operates without storage, the assessment focused on the potential impact of battery integration on self-consumption, self-sufficiency, and investment profitability under different energy-sharing schemes. The analysis considered a single community-scale battery rather than individual household storage systems, as larger shared installations are generally expected to benefit from economies of scale and a more efficient utilization of storage capacity.

We would like to acknowledge a limitation of this study: while in previous section we started from an optimization of the plant for the individual demands aggregated, the results presented are for a collective plant of 110,5kWp, that optimized by oemof. However here we have assumed the reality of having a PV plant of 116,05KWp. Very slight discrepancies may occur with the figures presented in the preceding section, but these minor variations, however, do not alter the overarching findings or the conclusions drawn from this work.

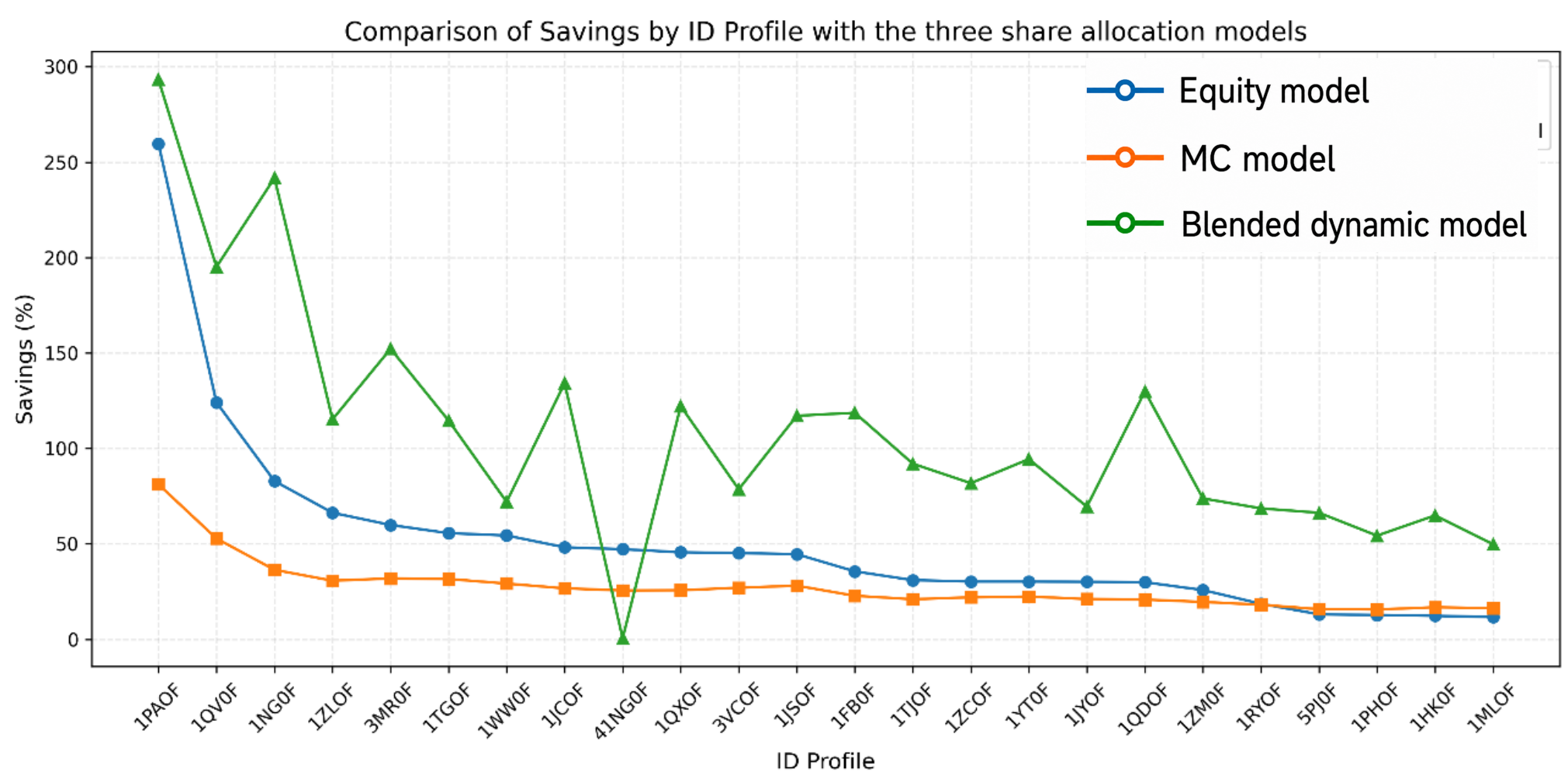

*Figure 2. Savings for each model compared to the same energy allocation strategy without battery. Savings are expressed as a percentage of the baseline annual cost, which explains why the trend is inversely related to absolute demand: profiles with lower consumption often achieve the highest relative savings, while high-demand profiles show lower percentages despite larger absolute savings in euros.*

From an economic perspective, the dynamic allocation strategy also provides the highest returns on battery investment, although in high-demand profiles such as 1PH0F (54,884 kWh/yr), 1ML0F (40,186 kWh/yr), and 1HK0F (20,704 kWh/yr), percentage savings are smaller because the battery offsets a smaller fraction of their large baseline costs, even though the euro savings are among the highest in absolute terms. The equity model yields moderate, more uniform percentages across profiles, typically between 20 % and 50 %, while the MC model consistently achieves the lowest relative savings, particularly for high-demand users. Overall, the results highlight that percentage savings are strongly influenced by demand size, and that dynamic allocation in its blended model maximises relative benefit across all demand levels, especially where storage flexibility can meet a large share of annual consumption.

Sensitivity analyses showed in Figure 3 and Figure 4 indicate that battery profitability increases with electricity prices and decreases with battery costs for all allocation methods. Nevertheless, dynamic allocation in its blended version remains economically attractive across a wider range of market conditions, while equity-based and MC-based approaches require higher electricity prices or lower storage costs to achieve comparable returns. Under current market conditions, these latter options are unlikely to provide sufficient economic incentives for citizens to invest in collective storage. The results therefore indicate that the viability of community-scale batteries depends not only on technology costs but also on the adoption of allocation mechanisms capable of maximizing the value of shared energy resources.

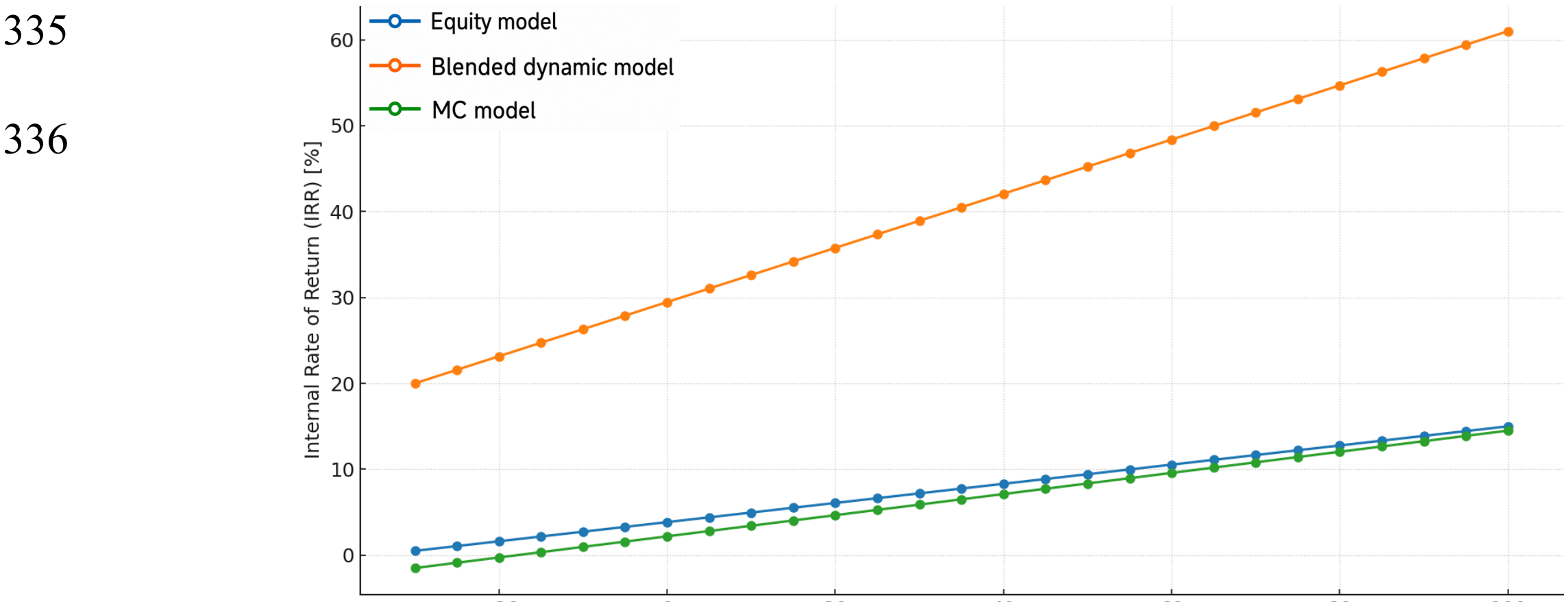

*Figure 3. Battery IRR vs. Electricity Price (energy).*

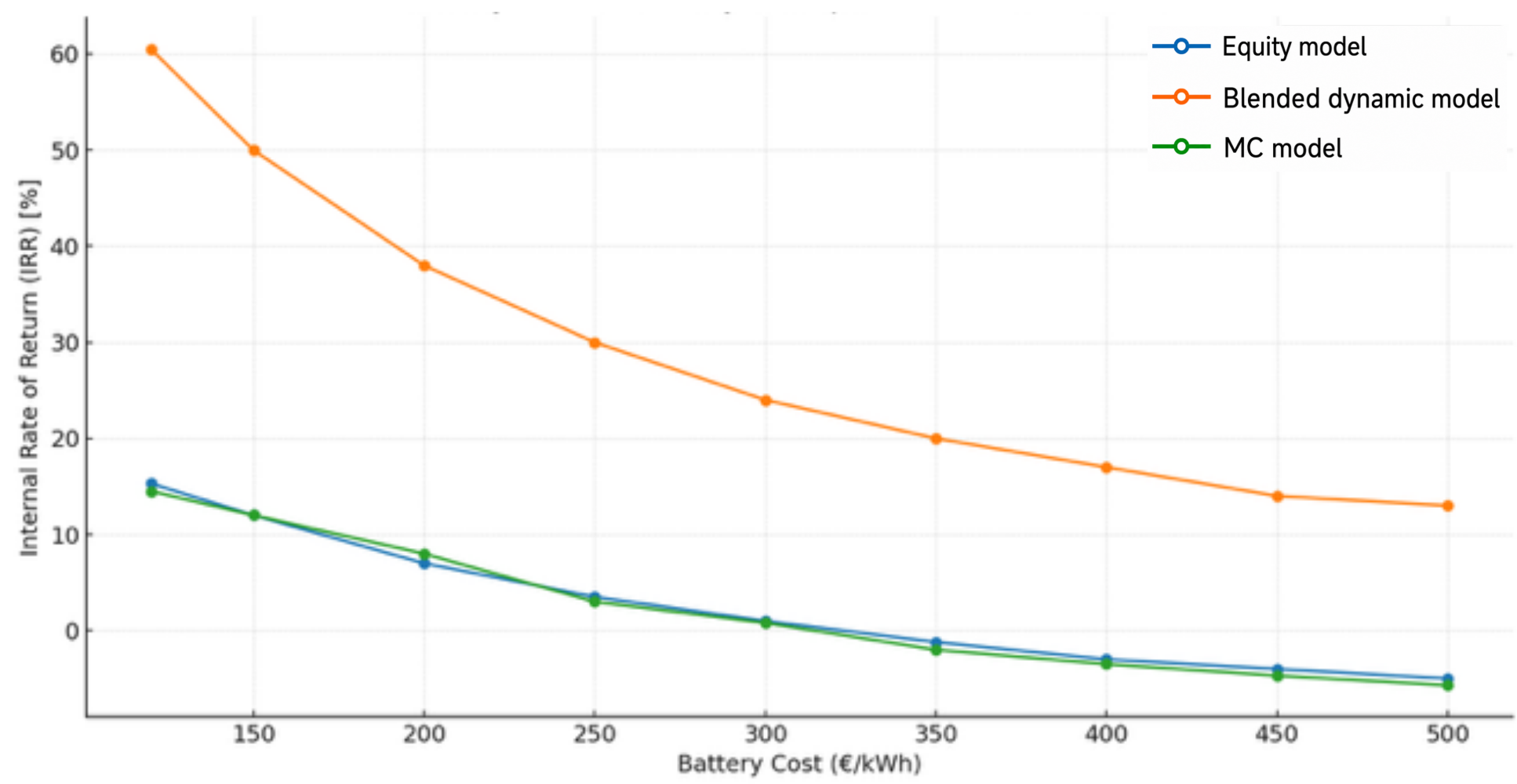

*Figure 4. Battery IRR vs. Battery Cost per KWh*

## 4. Recommendations

### 4.1 Policy recommendations

According to the study in this section, we aim at answering how self-consumption policies should be modified under the current low energy prices scenario.

The comparative exercise invites reflection on the role of public support mechanisms. At present, many EU member-states channels most renewable-energy subsidies to households through direct capital grants or tax rebates calculated on a € per-kWp basis. Under the individual scenario, these incentives are often essential: a 1 kWp rooftop system that costs €1,750 and saves roughly €65 per year (Table 3) would require more than twenty-five years to recover its investment without a grant, a period that exceeds typical investment lifetimes and erodes the project's internal rate of return. Consequently, small households tend to depend on subsidies to make individual self-consumption attractive.

The collective alternative paints a different picture. Even with no direct subsidy, the community plant achieves a *LCOE* of €0.023 $kWh^{-1}$ and cuts the variable portion of the electricity bill by fifty-five per cent, results that translate into a simple pay-back of barely five years and an after-tax internal rate of return close to 15 %. At those cost and benefit levels, an upfront grant would accelerate pay-back only marginally; the economic barrier becomes organisational rather than financial. It must be also highlighted that energy communities are not just a group of citizens but a legal entity, and this implies that other important cost that usually is charged to individual facilities, VAT, can be deducted when the facility is set by the energy community.

That is, energy communities are not dependant of subsides to be attractive. This observation suggests a shift in the optimal use of public funds. Rather than subsidising hardware that is already competitive at scale, policy could yield higher social returns by supporting the formation and professionalisation of energy communities without limiting their citizen-grassroots. Targeted measures might include seed funding for legal and technical advisory services, low-interest revolving loans for cooperatives, guarantees that reduce the cost of project finance and technical consultancy to implement optimization strategies such us those commented in this paper through better shares. Such instruments tackle transaction costs -permits, governance, metering integration- that individual subsidies leave untouched but that often determine whether a collective project ever reaches financial close.

Redirecting support in this manner would produce two complementary benefits. First, it would stretch limited public budgets further, because each euro spent on community facilitation leverages private capital at a larger scale than a traditional individual rooftop rebate. Second, it would enhance equity: low-income households that cannot pre-finance even a subsidised rooftop system could still join a cooperative by subscribing to a modest number of shares or by paying through their electricity bill, thereby gaining access to the same cost savings and to the governance rights enshrined in EU directives on citizen energy communities.

### 4.2 Allocation models recommendations

Table 7 highlights the trade-offs connected to the various allocation methods examined in this research. The findings indicate that there isn't one model that is the best for all situations, as the choice depends on the desired balance between technical effectiveness, cost savings, fairness, and ease of governance.

For communities mainly focused on boosting self-consumption and profits, flexible allocation based on past demand yields the best technical and financial results. However, the treatment of surplus energy remains a critical limitation. Since ownership of surplus generation is not explicitly defined, the practical implementation of this approach may face regulatory and governance challenges, particularly in jurisdictions where energy communities have limited capacity to manage or commercialise excess electricity.

For mature energy communities, the flexible model with surplus sharing seems to provide the most balanced approach. It keeps most of the effectiveness and savings from flexible allocation while ensuring a clear connection between investments and the distribution of benefits.

The idea of setting shares based on past hourly usage might be seen as a risk by members of the community. In Spain, users currently have access to only two years of past usage data, and yearly changes of up to 9,3% were noted in this study. As a result, allocation figures based on previous habits cannot ensure that future usage patterns will be accurately reflected. However, this limitation is somewhat eased by the rules, which permit hourly allocation figures shared with the DSO to be updated up to four times a year. Thus, historical-demand allocation should not be seen as a fixed rule, but as a flexible system that can change as usage patterns evolve.

Like the idea of a Typical Meteorological Year used in PV modelling, future applications could rely on typical demand profiles created from longer usage records, lessening the impact of unusual years and enhancing the strength of the allocation process.

The MCP approach provides an attractive mechanism for rewarding value creation and operational contribution. However, technical and economic optimality should not be confused with social optimality. By explicitly linking benefits to each member's contribution to collective profitability, MCP may implicitly classify some participants as net contributors and others as net beneficiaries. Although economically efficient, such differentiation may conflict with the cooperative principles that underpin many citizen-led energy communities and may reduce social acceptance if not accompanied by transparent governance and participatory decision-making processes.

Under these conditions, the hybrid 50% Equity–50% MCP scheme emerges as an attractive transition strategy for established communities seeking to move progressively from investment-based allocation towards performance-based mechanisms. By combining investment recognition with incentives linked to value creation, it mitigates many of the governance challenges associated with pure MCP schemes while improving the efficiency of energy distribution.

421

| Guiding principle | Allocation model | Technical performance (SCR/SSR) | Savings potential | Fairness perception | Main advantages | Main disadvantages | Governance complexity |
|---|---|---|---|---|---|---|---|
| **Prioritise investor confidence and administrative simplicity** | Fixed shares based on equity | **Medium** (does not adapt to hourly demand, leading to unused allocations and lower PV utilisation) | **Medium** (improves over individual systems but leaves part of the coordination value unexploited) | **High** (benefits remain proportional to financial contribution) | Simple, transparent and easy to communicate. | Limits the value created by load diversity. | Low |
| **Prioritise operational efficiency and value creation** | Dynamic* based on historical demand | **High** (maximises matching between generation and demand) | **High** (captures most of the available coordination value) | **Medium** (benefits may diverge from investment shares) | Maximises energy utilisation and savings. | Requires periodic recalculation and greater data availability. None is responsible for the surpluses => external consumers. | Medium-High |
| **Balance efficiency, savings and social acceptance** | Dynamic* + surplus redistribution | **High** (retains most efficiency gains of dynamic allocation) | **High** (preserves most community-wide savings) | **High** (maintains recognition of investment) | Balances efficiency and fairness. | Requires explicit surplus redistribution rules. | Medium |
| **Reward value creation rather than capital contribution** | MCP (Marginal Contribution to Profit) | **High** (rewards users that contribute most to collective optimisation) | **Medium–High** (captures substantial optimisation gains) | **Low–Medium** (may disadvantage large investors) | Strong incentive for efficient behaviour. | May undermine perceptions of solidarity and fairness by explicitly differentiating between members according to their | High |

| Guiding principle | Allocation model | Technical performance (SCR/SSR) | Savings potential | Fairness perception | Main advantages | Main disadvantages | Governance complexity |
|---|---|---|---|---|---|---|---|
| | | | | | | contribution to collective profitability. | |
| **Facilitate a gradual transition from investment-based to performance-based allocation** | Hybrid 50% Equity + 50% MCP | **Medium–High** (improves efficiency while retaining part of the equity logic) | **Medium–High** (captures part of the optimisation gains) | **High** (balances investor protection and operational contribution) | Provides a socially acceptable pathway for mature communities seeking to revise existing agreements without abandoning investment recognition. Mitigates many of the governance challenges associated with pure MCP schemes. | Does not fully maximise efficiency nor fully preserve investment proportionality. | Medium |

422 * Although not currently permitted by regulation, our proposal to implement this system relies on past consumption and generation data to predict future behaviour. This
423 information would be transparently communicated to the DSO, while surplus energy would be managed by the energy community itself, acting as an energy retailer

424 *Table 7. Comparison of the different schemes for an energy community aiming to re-allocate their shar*

### 4.3 Battery investments recommendations

The findings show that battery storage should not be seen as a one-size-fits-all answer for established energy communities that work under the present electricity market situation. While storage boosts the use of locally produced solar energy and lessens reliance on the grid, its financial appeal heavily depends on the energy-sharing system chosen by the community.

Blended dynamic allocation strategy regularly produced better financial returns and wider profit ranges compared to equity-based or MCP-based methods. With today's battery prices and electricity rates, investments in storage seem sensible mainly when allocation systems enhance the value of shared energy resources and enable the battery to adjust to real usage patterns. On the other hand, with fixed allocation systems, the extra value brought in by storage might not be enough to cover the cost of investment.

Thus, energy communities should focus on enhancing energy-sharing setups before making large investments in battery storage. Storage should be considered a later-stage optimization step instead of a requirement for boosting community performance.

## 5. Conclusions

This study investigated how citizen-led photovoltaic initiatives can maintain and enhance their value under increasingly challenging market conditions characterized by declining electricity prices and limited remuneration of surplus generation.

The first and most important finding is that collective self-consumption substantially outperforms equivalent individual photovoltaic installations. By aggregating diverse demand profiles, collective schemes require lower installed capacity, increase the local utilization of photovoltaic generation, and generate greater economic savings. These results suggest that, under current market conditions, the long-term development of citizen participation in photovoltaic self-consumption is likely to rely more on collective approaches than on fragmented individual installations.

Having established the advantages of collective self-consumption, the analysis explored which strategies can further improve the performance of energy communities. The results show that significant gains can still be achieved without additional investments. More adaptive allocation mechanisms increase the utilization of locally generated energy and improve overall community performance, making them an attractive low-cost strategy for existing initiatives. The study discusses on those that are likely to be implemented right now instead of presenting optimization strategies that in practice are not managerial.

Battery storage provides an additional source of flexibility and can further increase the value captured from photovoltaic generation. However, its benefits depend strongly on how energy is allocated and managed within the community. The results indicate that storage is most effective when combined with allocation mechanisms capable of maximizing the utilization of shared energy resources, suggesting that operational optimization should generally precede major investments in storage.

Overall, the findings indicate that the main challenge is no longer how to deploy photovoltaic capacity, but how to organize and manage it collectively. As renewable penetration continues to increase, and surplus electricity loses value, coordination, adaptive governance arrangements, and operational flexibility may become as important as generation technologies themselves in ensuring the long-term sustainability of citizen-led energy initiatives.

From a policy perspective, the results suggest that support schemes should not focus exclusively on promoting new generation capacity. Equal attention should be given to mechanisms that strengthen collective self-consumption, improve energy-sharing arrangements, and facilitate the efficient use of flexibility resources. These measures may prove essential for maintaining the economic attractiveness and social relevance of citizen participation in the energy transition.

## 6. Declarations

### a. Authors' Contributions

Individual contributions included the following: Conceptualization, AB. Cristobal; methodology, A.B. Cristóbal.; users data provision, L.M. Carrasco; formal analysis and investigation, A.B. Cristóbal, D. Sierra, L. Palomino; software, D. Sierra, L. Palomino; data curation and interpretation, A.B. Cristóbal and L. Narvarte; writing—original draft preparation, A.B. Cristóbal; writing—tables and graphs preparation, , D. Sierra. L. Palomino; writing—review and editing, AB. Cristóbal, L. Narvarte; funding acquisition, A.B. Cristóbal. All authors have read and agreed to the published version of the manuscript.

### b. Data availability

The datasets generated during and/or analysed during the current study are available from the corresponding author on reasonable request.

### c. Competing interests

The authors have no financial or proprietary interests in any material discussed in this article.

### d. AI tools

During the preparation of this work, the author(s) used GEMINI PRO licensed to Universidad Politécnica de Madrid for improving readability and finding typos. The author(s) reviewed and edited the output as needed and take full responsibility for the content of the published article.

## 7. Acknowledgements

This research has been funded by the project "FOREVERPV-CM: For an Environmentally Friendly Photovoltaic Technology" (TEC2024/ECO72), granted by the Comunidad de Madrid. Data and experience are coming from the European project LIFE-ENERCOM-JALON-101076395.